\documentclass[a4paper,fleqn,usenatbib]{mnras}

\usepackage{newtxtext,newtxmath}

\usepackage[T1]{fontenc}
\usepackage{ae,aecompl}

\usepackage{graphicx}	
\usepackage{amsmath}	
\usepackage{amssymb}	

\def\vector#1{\mbox{\boldmath $#1$}}

\title[Suppression of recycling by buoyancy barrier]{Suppression of atmospheric recycling of planets embedded in a protoplanetary disc by buoyancy barrier}

\author[H. Kurokawa and T. Tanigawa]{
Hiroyuki Kurokawa,$^{1}$\thanks{E-mail: hiro.kurokawa@elsi.jp (HK)}
Takayuki Tanigawa$^{2}$
\\
$^{1}$Earth-Life Science Institute, Tokyo Institute of Technology, 2-12-1 Ookayama, Meguro, Tokyo 152-8550, Japan\\
$^{2}$National Institute of Technology, Ichinoseki College, Takanashi, Hagisho, Ichinoseki-shi, Iwate 021-8511, Japan
}

\date{Accepted XXX. Received YYY; in original form ZZZ}

\pubyear{2017}

\begin{document}
\label{firstpage}
\pagerange{\pageref{firstpage}--\pageref{lastpage}}
\maketitle

\begin{abstract}
The ubiquity of super-Earths poses a problem for planet formation theory to explain how they avoided becoming gas giants.
Rapid recycling of the envelope gas of planets embedded in a protoplanetary disc has been proposed to delay the cooling and following accretion of disc gas.
We compare isothermal and non-isothermal 3D hydrodynamical simulations of the gas flow past a planet to investigate the influence on the feasibility of the recycling mechanism.
Radiative cooling is implemented by using the $\beta$ cooling model.
We find that, in either case, gas enters the Bondi sphere at high latitudes and leaves through the midplane regions, or vice versa when disc gas rotates sub-Keplerian.
However, in contrast to the isothermal case where the recycling flow reaches \textcolor{black}{the deeper part of the envelope}, 
the inflow is inhibited from reaching the \textcolor{black}{deep envelope} in the non-isothermal case.
Once the atmosphere starts cooling, buoyant force prevents the high-entropy disc gas from intruding the low-entropy atmosphere.
We suggest that the buoyancy barrier isolates the lower envelope from the recycling and allows further cooling, which may lead runaway gas accretion onto the core.
\end{abstract}

\begin{keywords}
hydrodynamics -- planets and satellites: atmospheres -- planets and satellites: formation -- protoplanetary discs.
\end{keywords}



\section{Introduction} \label{sec:intro}

The {\it Kepler} mission has revealed that about half of Sun-like stars harbour super-Earths, here defined as planets whose radii are 1--4 $R_{\earth}$, with periods $<$ 85 days \citep[e.g.,][]{Fressin+2013}.
Those super-Earths have the masses of 2--20 $M_{\earth}$ \citep[e.g.,][]{Weiss+Marcy2014} and their mass-radius relations indicate that the gas-to-core mass ratios are $\sim$1--10\% \citep[e.g.,][]{Lopez+Fortney2014}.
In contrast, giant planets are rare in the close-in orbits: the occurrence is $\sim$1\% \citep[e.g.,][]{Fressin+2013}.

The ubiquity of super-Earths poses a problem for planet formation theory: how did they avoid becoming gas giants?
A solid core embedded in a protoplanetary disc captures the surrounding disc gas within the Bondi or Hill spheres.
Radiative cooling and contraction of the envelope gas allows further accretion of disc gas \citep[e.g.,][]{Lee+2014}. 
If the mass of the core exceeds 5--20 $M_{\earth}$, it undergoes runaway gas accretion to become a gas giant \citep[e.g.,][]{Mizuno1980,Pollack+1996,Ikoma+2000}.
The critical core mass becomes even smaller if the envelope was polluted with heavy elements \citep[][]{Hori+Ikoma2011,Venturini+2015}.

Late-stage core formation has been proposed to solve the problem of ubiquitous super-Earths.
Given the short time-scale of the runaway gas accretion ($\sim$1 Myr for a 10 $M_{\earth}$ core),  \textcolor{black}{it has been} proposed that super-Earth cores coagulated just as \textcolor{black}{disc} gas was about to disappear completely \textcolor{black}{so that they did not have sufficient time to undergo runaway accretion \citep{Ikoma+Hori2012,Lee+2014}}.
The final assembly during disc dispersal is the expected result in conventional planet formation theory due to mutual gravitational interactions of proto-cores \citep[e.g.,][]{Kominami+Ida2002,Inamdar+Schlichting2015}.

The feasibility of the late-stage core formation scenario depends on the planet formation regimes.
The classical planet formation theory assumed that the proto-cores are formed by planetesimal accretion.
The isolation mass of the planetesimal accretion is small at $\sim 0.1\ {\rm AU}$ \citep{Kokubo+Ida2000}, which is enough to avoid the runaway gas accretion before the final assembly.
On the other hand, in the pebble accretion theory \citep{Ormel+Klahr2010,Lambrechts+Johansen2012}, where proto-cores accrete pebbles drifting from the outer regions of the disc, proto-cores can grow up to a mass heavy enough to carve a gap in the pebble disc \citep{Lambrechts+2014}.
The inward drift of pebbles has been suggested observationally from the disc sizes, which are found to be more compact in the continuum (pebbles) than in the lines (gas) \citep[][and references therein]{Ormel2017}, and from dust rings, which might be formed by the drift and sintering of icy dust \citep{Okuzumi+2016}.

Other mechanisms have been suggested to inhibit super-Earths from becoming gas giants by delaying the cooling and contraction of their proto-envelopes.
When super-Earths are on eccentric orbits, stellar tidal heating would suppress the cooling \citep{Ginzberg+Sari2017}.
\citet{Yu2017} has proposed that the tidally-forced turbulent diffusion may affect the heat transport inside the planet and may explain the fact that super-Earths are commonly found in the inner protoplanetary disc.

Another mechanism to explain the ubiquity of super-Earths utilises a hydrodynamical explanation -- a rapid recycling of the atmospheric gas embedded in a protoplanetary disc may prevent the envelope from cooling \citep{Ormel+2015}.
They conducted isothermal hydrodynamical simulations of flow past a planet embedded in a protoplanetary disc.
The atmosphere inside the Bondi radius was shown to be an open system where disc gas enters from high latitude (inflow) and leaves through the midplane region (outflow) in cases where shear flow was adopted.
They also found that the topography changes when the rotation of disc gas is sub-Keplerian, but the recycling was observed in any cases.
They argued that the recycling is faster than the cooling of the envelope gas for close-in super-Earths, and so that further accretion of disc gas is prevented.

The recycling mechanism works only if the high-entropy disc gas can replace the low-entropy atmospheric gas.
The treatment of the thermal structure is critical to assess whether the recycling can halt the cooling and further accretion of disc gas.
\textcolor{black}{Rapid recycling has been observed in isothermal simulations \citep{Ormel+2015,Fung+2015}.
Adiabatic simulations have also found no dynamical boundary between the atmosphere and disc gas \citep{Popovas+2018}.
In contrast,} non-isothermal, radiative-hydrodynamical simulations have found a bound atmospheric region around a planet below the recycling region \citep{DAngelo+Bodenheimer2013,Cimerman+2017,Lambrechts+Lega2017}.
Why the efficiency of recycling differs between isothermal and non-isothermal models is poorly understood, and is the focus of this study.

We conduct a comparison between isothermal and non-isothermal 3D hydrodynamical simulations of the flow around a planet embedded in a protoplanetary disc to investigate the effect of thermal profiles on the feasibility of the recycling mechanism.
Section \ref{sec:model} presents the model.
Section \ref{sec:results} shows the results.
The implications for the formation of super-Earths are discussed in Section \ref{sec:discussion}.
We conclude in Section \ref{sec:conclusion}.

\section{Model} \label{sec:model}

\begin{table*}
    \centering
    \begin{tabular}{lllllll}
    \hline
    Name & Cooling time & Mass & Domain size & Headwind speed & Injection time &  Termination time \\
     & $\beta$ & $m$ & $r_{\rm out}$ & $M_{\rm hw}$ & $t_{\rm inj}$ & $t_{\rm end}$ \\
    \hline
    {\tt shear-is-m001} & isothermal & 0.01 & 0.5 & 0 & 0.5 & 15 \\
    {\tt shear-B4-m001} & $10^{-4}$ & 0.01 & 0.5 & 0 & 0.5 & 15 \\
    {\tt shear-B3-m001} & $10^{-3}$ & 0.01 & 0.5 & 0 & 0.5 & 15 \\
    {\tt shear-B2-m001} & $10^{-2}$ & 0.01 & 0.5 & 0 & 0.5 & 15 \\
    {\tt shear-ad-m001} & adiabatic & 0.01 & 0.5 & 0 & 0.5 & 15 \\
    {\tt shear-is-m01} & isothermal & 0.1 & 5 & 0 & 0.5 & 15 \\
    {\tt shear-B2-m01} & $10^{-2}$ & 0.1 & 5 & 0 & 0.5 & 15 \\
    {\tt shear-is-m1} & isothermal & 1 & 5 & 0 & 1 & 30 \\
    {\tt shear-B2-m1} & $10^{-2}$ & 1 & 5 & 0 & 1 & 30 \\
    {\tt shear-B0-m1} & 1 & 1 & 5 & 0 & 1 & 30 \\
    {\tt headw-is-m001} & isothermal & 0.01 & 0.5 & 0.1 & 0.5 & 10 \\
    {\tt headw-B2-m001} & $10^{-2}$ & 0.01 & 0.5 & 0.1 & 0.5 & 10 \\
    {\tt heado-is-m001} & isothermal & 0.01 & 0.5 & 0.1 & 0.5 & 10 \\
    {\tt heado-B2-m001} & $10^{-2}$ & 0.01 & 0.5 & 0.1 & 0.5 & 10 \\
    \hline
    \end{tabular}
    \caption{List of the simulations. The first column gives the name where {\tt shear}, {\tt headw}, and {\tt heado} depict shear flow, shear flow with headwind, and headwind only, respectively. \textcolor{black}{The second to seventh columns show values of the cooling time, planetary mass, computational domain size, headwind speed, injection time of gravity of the planet, and termination time of computation, respectively.}}
    \label{tab:models}
\end{table*}

Flow around a planet embedded in a protoplanetary disc was calculated in this study.
We followed the setup of the hydrodynamical simulations of \citet{Ormel+2015} except for \textcolor{black}{three} major differences.
i) Whereas \citet{Ormel+2015} conducted isothermal simulations, we performed both isothermal and non-isothermal simulations where the radiative cooling was implemented by using the $\beta$ cooling model (subsection \ref{subsec:equ}).
ii) \citet{Ormel+2015} implemented a modified gravitational potential of a planet to avoid numerical difficulties found in their study. 
We conducted simulations using a standard gravitational potential (subsection \ref{subsec:equ}).
\textcolor{black}{iii) Whereas \citet{Ormel+2015} omitted the vertical stratification of the circumstellar disc for simplification, we considered the cases both with and without the stratification, for lower and higher mass planets, respectively (subsection \ref{subsec:settings}).}

\subsection{Equations} \label{subsec:equ}

A compressible, inviscid fluid \textcolor{black}{of an ideal gas} was assumed in this study.
We used dimensionless units where velocities are expressed in terms of the sound speed $c_{\rm s}$, times in terms of the reciprocal of the Keplerian frequency $\Omega^{-1}$, and lengths in terms of the disc scale height $H \equiv c_{\rm s}/\Omega$ \textcolor{black}{of the background gas}.
The dimensionless mass of a planet $m$ follows the relation $m=R_{\rm Bondi}/H$, where $R_{\rm Bondi}$ is the Bondi radius.
\textcolor{black}{The Hill radius is given by $R_{\rm Hill} = (m/3)^{1/3}$.
As the gravity of the gas itself was neglected in this study, we can define another dimensionless unit to measure the densities.
Here, the background gas density at the midplane $\rho_{\rm 0}$ was set as unity.}

\textcolor{black}{
The scaling gives the planetary mass $M_{\rm p} = m c_{\rm s}^3/(G \Omega)$, where $G$ is the gravitational constant.
Assuming a solar-mass host star and a disc temperature profile $T = 270\ (a/1\ {\rm AU})^{-1/2}$ \citep[the minimum-mass solar nebula model,][]{Weidenschilling1977,Hayashi+1985}, $M_{\rm p}$ is given by, 
\begin{equation}
    M_{\rm p} \simeq 12\ m \biggl( \frac{a}{1\ {\rm AU}} \biggr)^{\frac{3}{4}} M_{\earth}, \label{eq:mass}
\end{equation}
where $a$ is the orbital radius. 
}

The continuity, Euler's, and energy conservation equations are given by,
\begin{equation}
    \frac{\partial \rho}{\partial t} + \nabla \cdot (\rho \vector{v}) = 0,
\end{equation}
\begin{equation}
    \frac{\partial }{\partial t} (\rho \vector{v}) + \nabla \cdot (\rho \vector{v} \vector{v}) = \nabla p + \rho (\vector{F}_{\rm cor} + \vector{F}_{\rm tid} + \vector{F}_{\rm hw} + \vector{F}_{\rm p}),
\end{equation}
\begin{equation}
    \begin{split}
        \frac{\partial E}{\partial t} + \nabla \cdot [(E+p)\vector{v}] =  \rho \vector{v} \cdot (\vector{F}_{\rm cor} + \vector{F}_{\rm tid} +\vector{F}_{\rm hw} + \vector{F}_{\rm p}) \\
        \quad -\frac{U(\rho,T)-U(\rho,T_{\rm 0})}{\beta}, \label{eq:energy}
    \end{split}
\end{equation}
where $\vector{v}$ is the velocity, $\rho$ is the density, $p$ is the pressure.
The internal and total energy densities $U$ and $E$ are given by,
\begin{equation}
    U = \frac{p}{\gamma-1},
\end{equation}
\begin{equation}
    E = U+\frac{1}{2}\rho v^2,
\end{equation}
where $\gamma$ is the ratio of specific heat.
\textcolor{black}{We assumed $\gamma=7/5$.}

The last term in Equation \ref{eq:energy} is the thermal relaxation approximated by the $\beta$ cooling model where the temperature $T$ relaxes towards the background temperature $T_0$ with the dimensionless time-scale $\beta$ \textcolor{black}{\citep[e.g.,][]{Gammie2001}}.
\textcolor{black}{The $\beta$ cooling was implemented in order to replicate the influence of radiative cooling on the recycling flow (Section \ref{sec:intro}).
As we will show in the following sections, our model is useful to understand the mechanism which determines the efficiency of atmospheric recycling.
The advantages to using the $\beta$ cooling model are, i) it is computationally less expensive than radiative-transfer calculations, and ii) we can obtain insights into the effect of radiative cooling by changing $\beta$ values artificially.
The disadvantages are, i) this approximation becomes unrealistic for optically thick and non-linear regimes, and ii) we assume one $\beta$ value even though the relaxation time may vary from place to place: the low-density disc gas and dense atmosphere have different relaxation time scales.
We will change $\beta$ values as a free parameter in Section \ref{sec:model} in order to understand the influence.
In subsection \ref{subsec:accretion}, we will discuss the implications for gas accretion onto planets by assuming that the $\beta$ value represents the relaxation time of disc gas descending into the Bondi radius.
}

The forces in the non-inertial frame read as follows: the Coriolis force $\vector{F}_{\rm cor} = -2 \vector{e}_z \times \vector{v}$, the tidal force $\vector{F}_{\rm tid} = 3x  \vector{e}_x \textcolor{black}{-z  \vector{e}_z}$ (where $\vector{e}_i$ is the unit vector in the $i$-direction), the global pressure force due to the sub-Keplerian motion of the gas $\vector{F}_{\rm hw} = 2M_{\rm hw} \vector{e}_x$ (where $M_{\rm hw}$ is the Mach number of the headwind), and the gravitational force due to the planet $\vector{F}_{\rm p}$.
\textcolor{black}{Here we assumed that the centre of the planet is located at the origin of the coordinates.
For the lowest mass simulations ($m=0.01$), the tidal force in the $z$-direction was neglected following \citet{Ormel+2015} (subsection \ref{subsec:settings}).}

The gravity term implementing a smoothing in space was gradually inserted over time:
\begin{equation}
    \vector{F}_{\rm p} = -\nabla \biggl( \frac{m}{\sqrt{r^2+\textcolor{black}{r_{\rm sm}^2}}}  \biggr) \cdot \biggl\{ 1-\exp\biggl[ -\frac{1}{2}\biggl(\frac{t}{t_{\rm inj}} \biggr)^2 \biggr] \biggr\}, \label{eq:Fp}
\end{equation}
where $m$ is the dimensionless mass of the planet, $r$ is the distance from the centre of the planet, \textcolor{black}{$r_{\rm sm}$ is the smoothing length, and $t_{\rm inj}$ is the injection time.
We assumed $r_{\rm sm} = 10^{-1} \times m$.}
We note that Equation \ref{eq:Fp} differs from the form used in \citet{Ormel+2015} because they adopted a modified gravitational potential of a planet to avoid numerical difficulties found in their study.

\subsection{Numerical settings} \label{subsec:settings}

We solved the equations described in subsection \ref{subsec:equ} by using the hydrodynamical simulation code Athena++  \citep[][Stone et al. in prep.]{White+2016}, an extended re-write of the Athena astrophysical magnetohydrodynamics code \citep{Stone+2008}.
We adopted a spherical grid with coordinates ($r, \theta, \phi$), centred on the planet.
In order to resolve flow inside the Bondi radius, a logarithmic grid for the radial dimension was employed.
We adopted polar mesh spacing proportional to $(3\psi^2 + 1)$ where $\psi$ is the angle from the midplane, which produces high resolution near the midplane.

The unperturbed solution to the shearing-sheet was adopted for the initial and outer-boundary conditions: 
\begin{equation}
    \vector{v} = \biggl( -\frac{3}{2}x - M_{\rm hw} \biggl) \vector{e}_y,\ \textcolor{black}{\rho = \exp{\biggl( -\frac{1}{2} z^2 \biggr) }},
\end{equation}
\textcolor{black}{where the density has vertical stratification due to the stellar gravity (the tidal force in the $z$-direction in our non-inertial frame).
In the case of the lowest mass planet ($m = 0.01$) where we omitted the $z$-direction tidal force, we adopted $\rho = \rho_{\rm 0} = 1$}.
The reflective boundary condition was assumed for the inner boundary.
However, after conducting several test runs\textcolor{black}{,} we found that there is unphysical energy flow at the inner boundary when gas accretes, which affected the entire calculated domain.
This seems to be caused by the reflective boundary condition as it is precisely reflective only for the (uniform) Cartesian grid.
In order to prevent this unphysical energy flux at the inner boundary from affecting the entire flow, we introduced an artificial cooling at the inner boundary where a small value of $\beta$ (\textcolor{black}{$= 10^{-5}$}) was adopted.

A summary of simulations is given in Table \ref{tab:models}.
We assumed the planetary mass $m = 0.01$ \textcolor{black}{as a nominal value following \citet{Ormel+2015}, which allows us to compare our results to theirs.
This dimensionless mass corresponds to a 0.12 $M_{\earth}$ planet orbiting a solar-mass star at 1 AU (Equation \ref{eq:mass}). 
The $z$-direction tidal force and vertical stratification were omitted in the nominal case.
For higher mass cases ($m = 0.1$ and $m = 1$, corresponding to 1.2 $M_{\earth}$ and 12 $M_{\earth}$ planet at 1 AU), the density stratification was considered. 
We note that, because our local simulations do not allow a planet to open a ringed gap in the disc, low mass planets ($m < 1$) would be preferred.
Nevertheless, high mass simulations ($m = 1$) were performed in order to check that our conclusions do not depend on the assumed planetary mass.}
The inner domain boundary was $r_{\rm inn}=10^{-3}$. 
The numerical resolution was $[r,\theta,\phi] = [128,64,128]$.

\section{Results} \label{sec:results}

\begin{figure*}
    \centering
    \includegraphics[width=\linewidth]{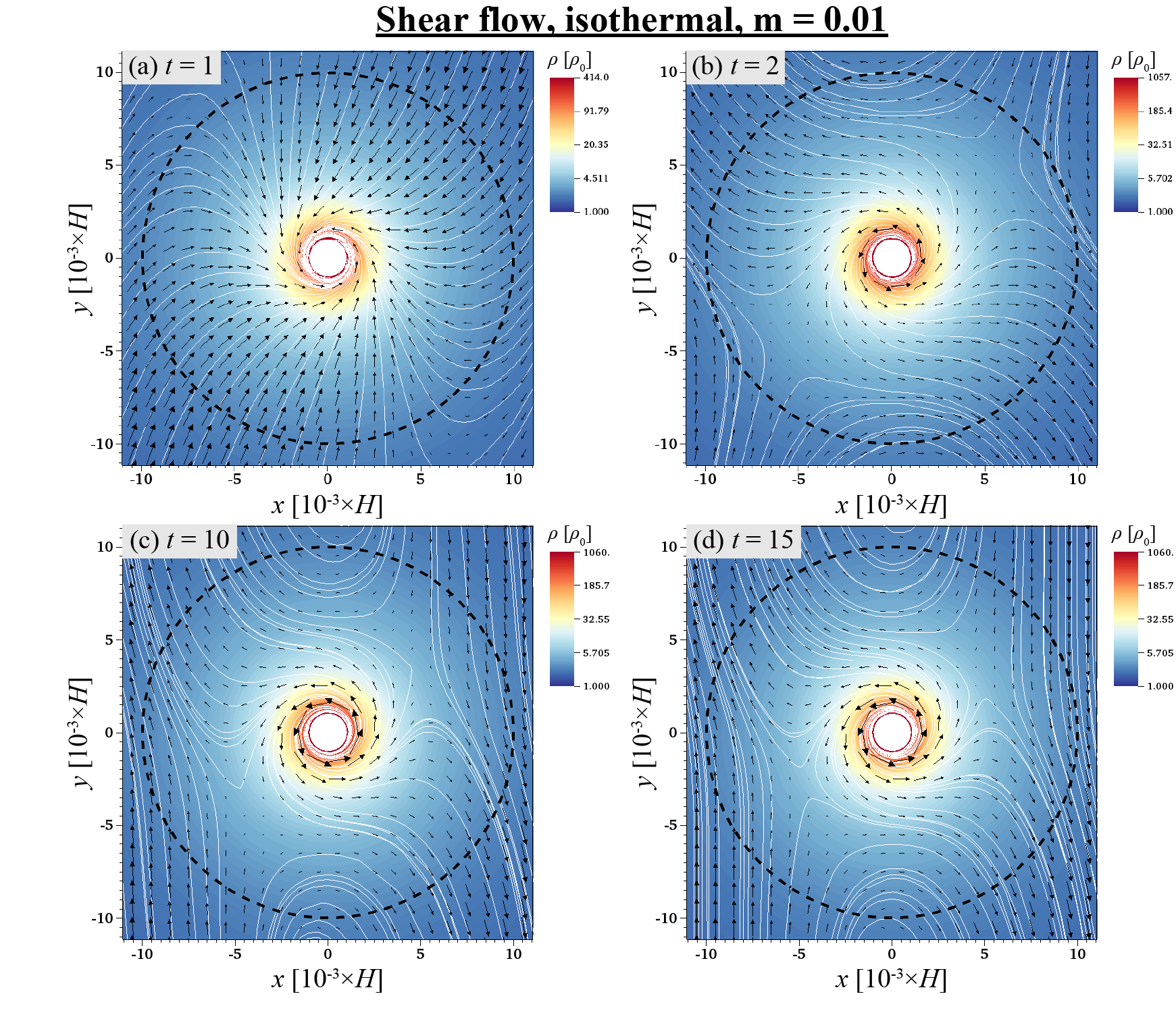}
    \caption{Midplane time sequence of {\tt shear-is-m001} run at $t=1, 2, 10$, and $15$. The flow pattern is represented by arrows and streamlines. \textcolor{black}{The length of the arrows scales to the speed.} The density is shown by colour contour. A dashed circle denotes the Bondi Radius.}
    \label{fig:time_iso}
\end{figure*}

\begin{figure*}
    \centering
    \includegraphics[width=\linewidth]{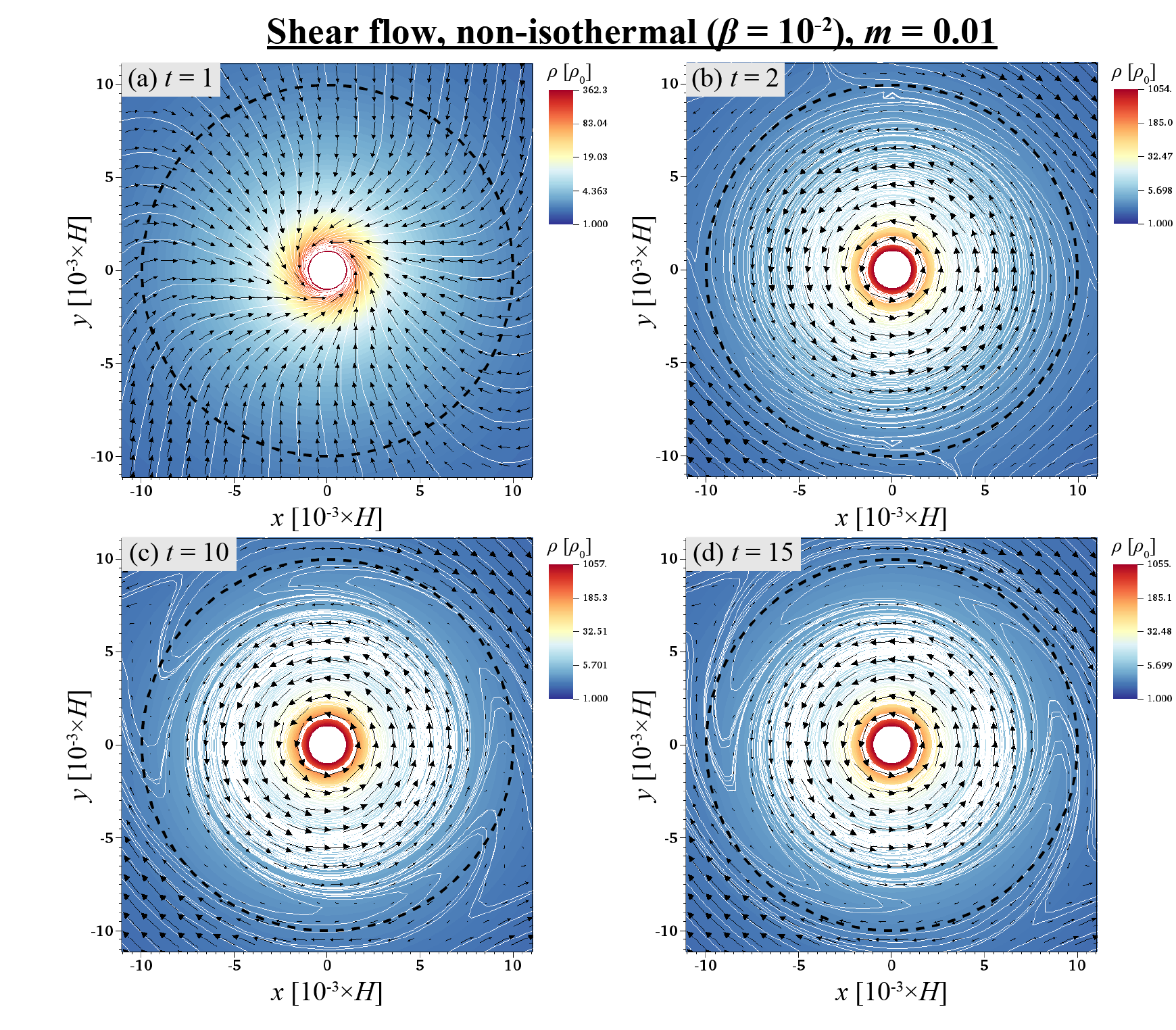}
    \caption{\textcolor{black}{The same as Figure \ref{fig:time_iso}, but the results of {\tt shear-B2-m001} run are shown.}}
    \label{fig:time_B2}
\end{figure*}

\begin{figure}
    \centering
    \includegraphics[width=\linewidth]{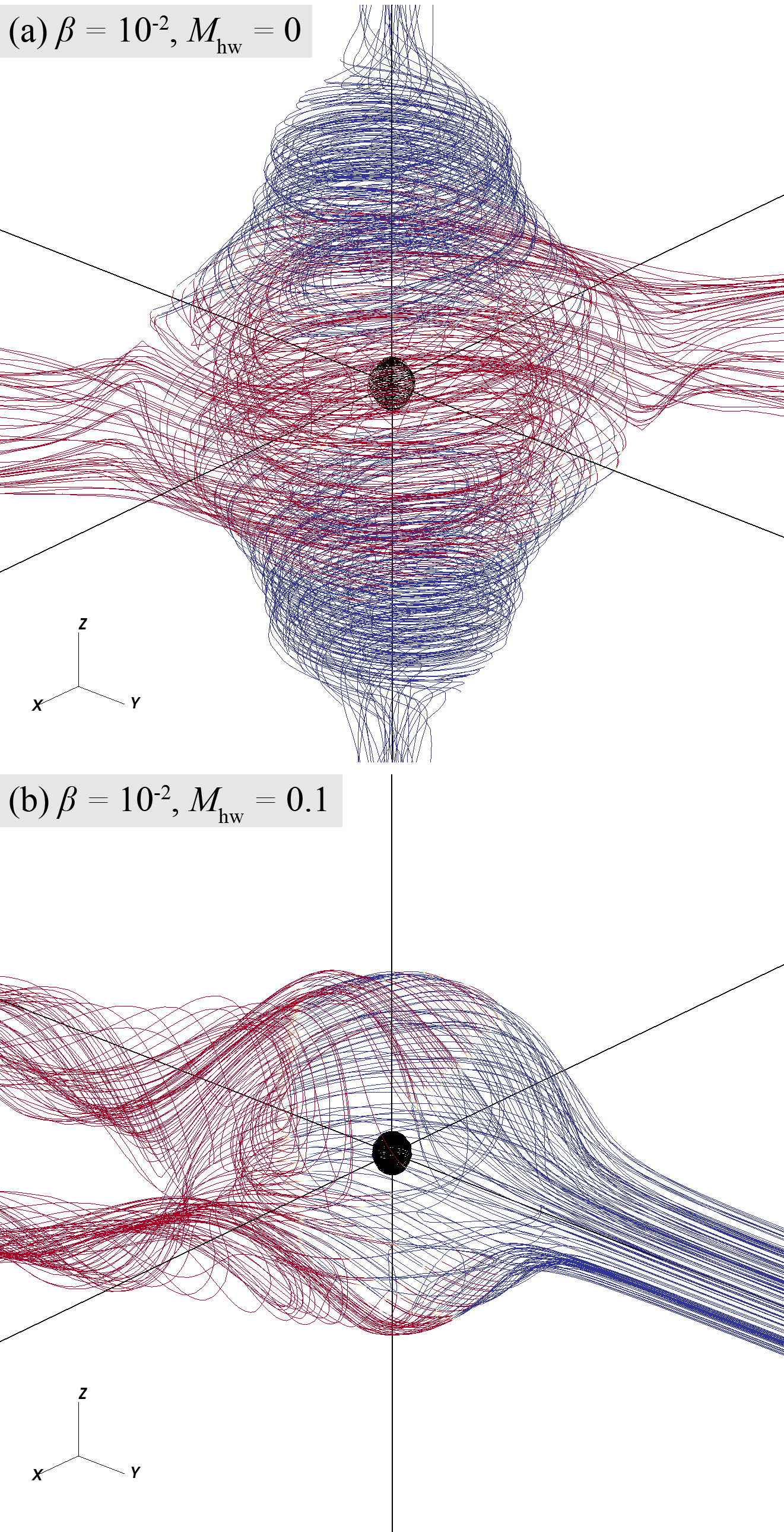}
    \caption{3D streamlines near the Bondi sphere in (a) {\tt shear-B2-m001} and (b) {\tt headw-B2-m001} runs at $t=5$. Colour represents where the radial speed is negative (blue) or positive (red). The inner black lines show the inner domain boundary.}
    \label{fig:3D}
\end{figure}

\begin{figure*}
    \centering
    \includegraphics[width=0.95\linewidth]{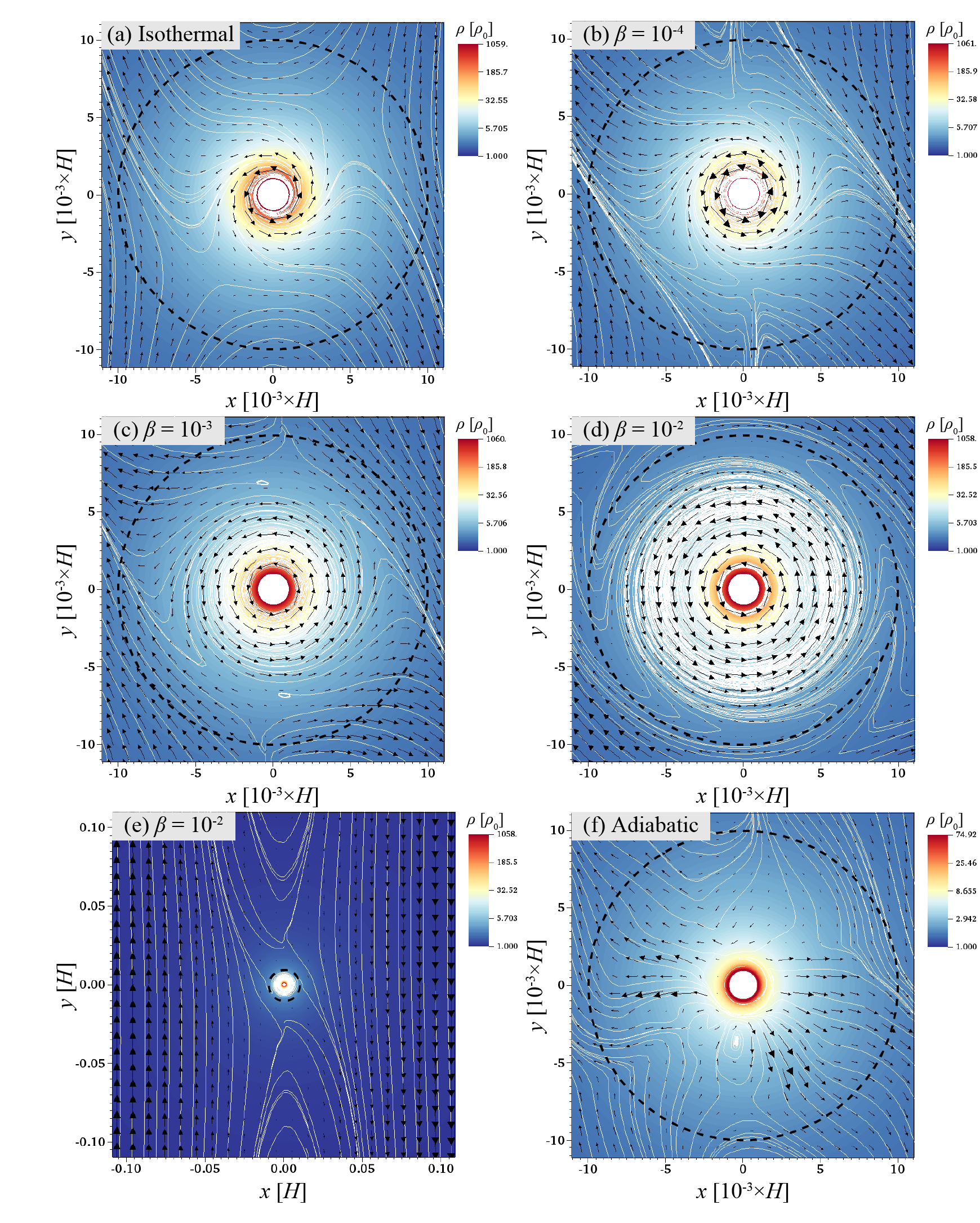}
    \caption{A comparison of flow patterns in the midplane at $t=7$: \textcolor{black}{(a) {\tt shear-is-m001}, (b) {\tt shear-B4-m001}, (c) {\tt shear-B3-m001}, (d) {\tt shear-B2-m001}, (e) the large-scale structure of {\tt shear-B2-m001}, and (f) {\tt shear-ad-m001}}. The flow pattern is represented by arrows and streamlines. \textcolor{black}{The length of the arrows scales to the speed, but the scale is different for each figures}. The density is shown by colour contour. A dashed circle denotes the Bondi radius.}
    \label{fig:comp_non-iso}
\end{figure*}

\begin{figure}
    \centering
    \includegraphics[width=\linewidth]{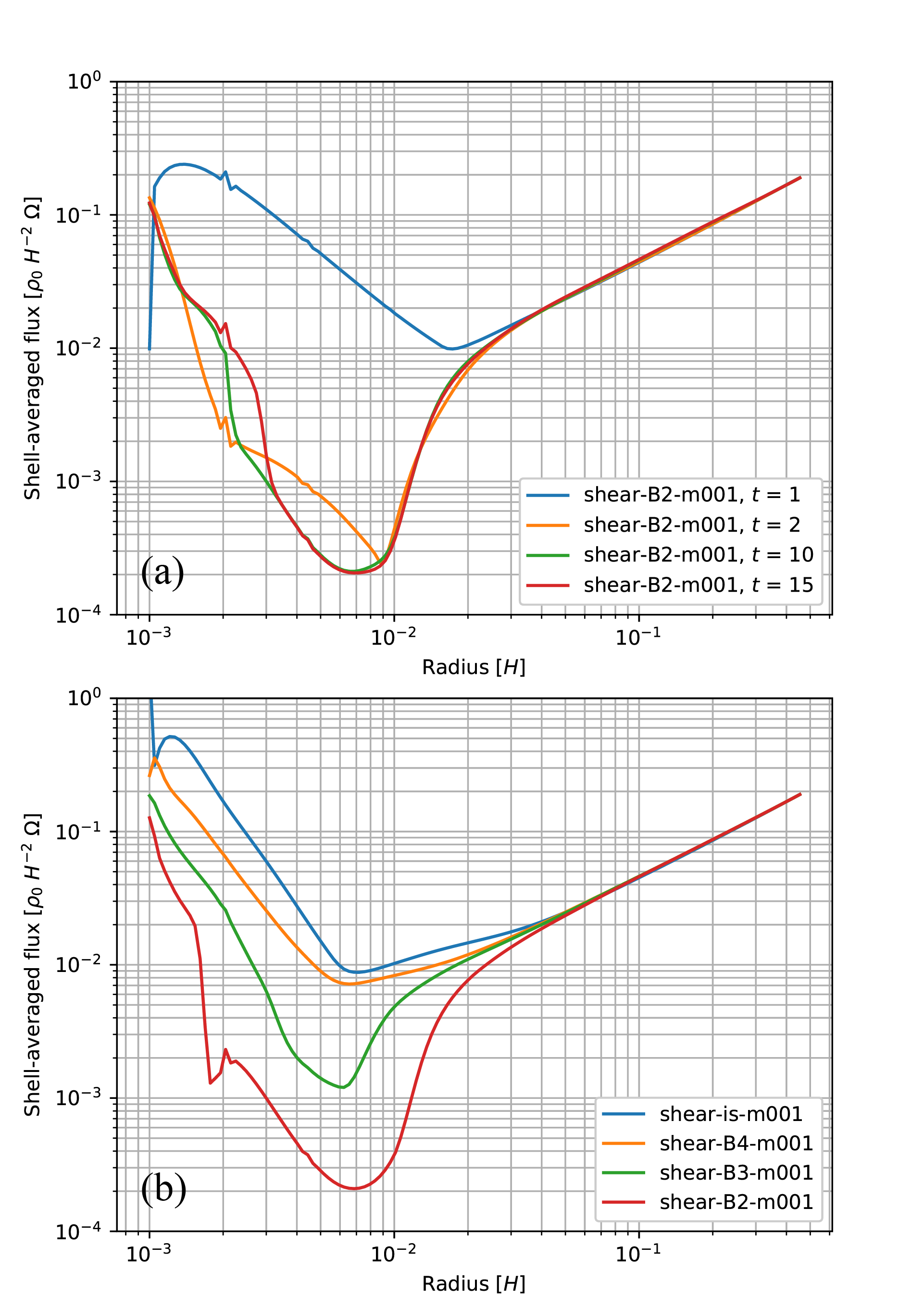}
    \caption{\textcolor{black}{Shell-averaged recycling flux (Equation \ref{eq:flux}). (a) Time sequence of {\tt shear-B2-m001} run and (b) a comparison between isothermal ({\tt shear-is-m001}) and non-isothermal ({\tt shear-B4-m001}, {\tt shear-B3-m001}, and {\tt shear-B2-m001}) runs at $t = 7$ are shown.}}
    \label{fig:r-flux}
\end{figure}

\begin{figure*}
    \centering
    \includegraphics[width=0.95\linewidth]{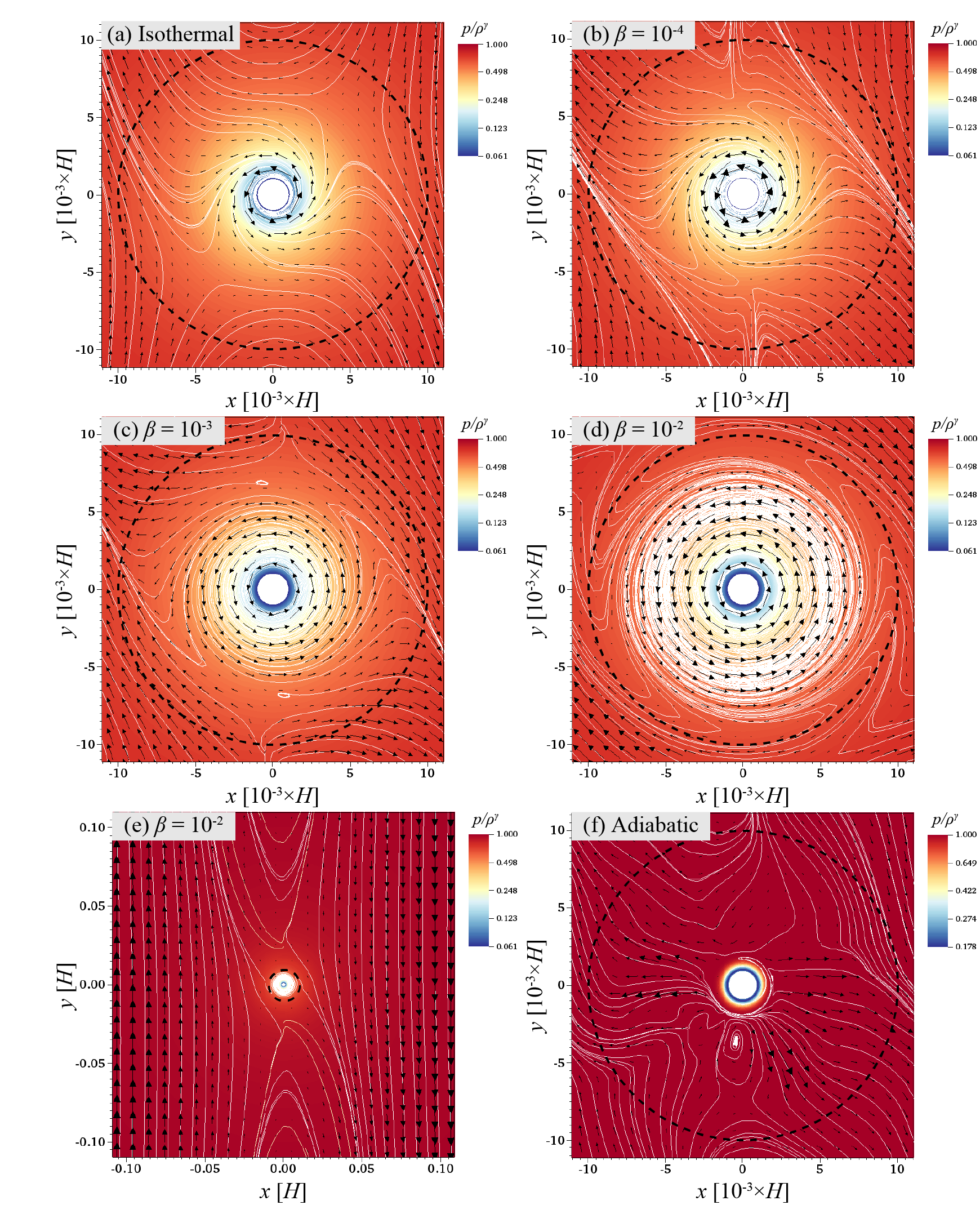}
    \caption{Data are the same as Figure \ref{fig:comp_non-iso}, but colour contour represents $p/\rho^\gamma$, which is the measure of entropy.}
    \label{fig:entropy}
\end{figure*}

\begin{figure*}
    \centering
    \includegraphics[width=\linewidth]{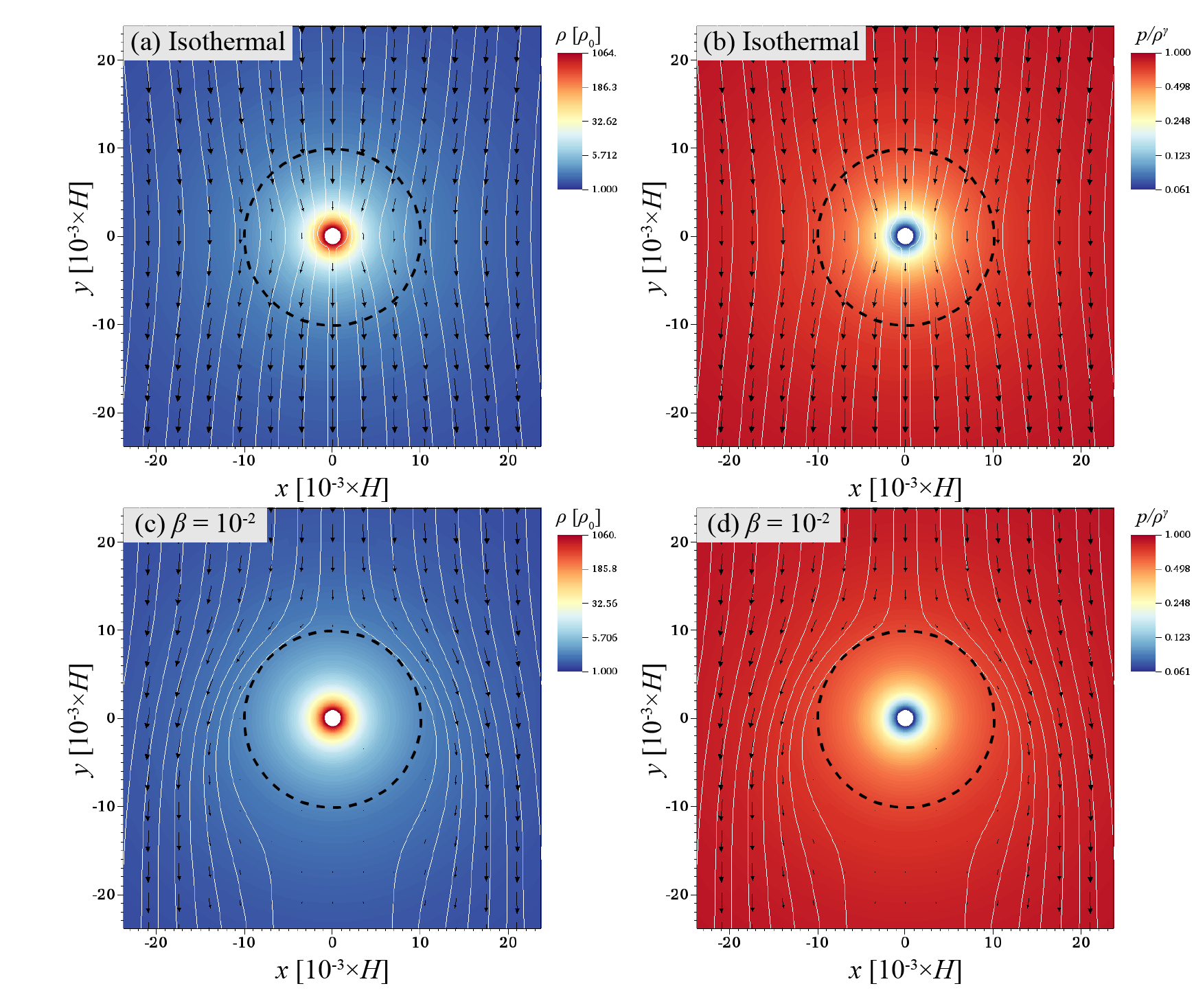}
    \caption{A comparison of flow patterns of (a,b) {\tt heado-is-m001} and (c,d) {\tt heado-B2-m001} (headwind only). The flow pattern is represented by arrows and streamlines. \textcolor{black}{The length of the arrows scales to the speed.} Colour contour represents (a,c) density and (b,d) $p/\rho^\gamma$ (entropy), respectively. A dashed circle denotes the Bondi Radius.}
    \label{fig:headonly}
\end{figure*}

\begin{figure*}
    \centering
    \includegraphics[width=\linewidth]{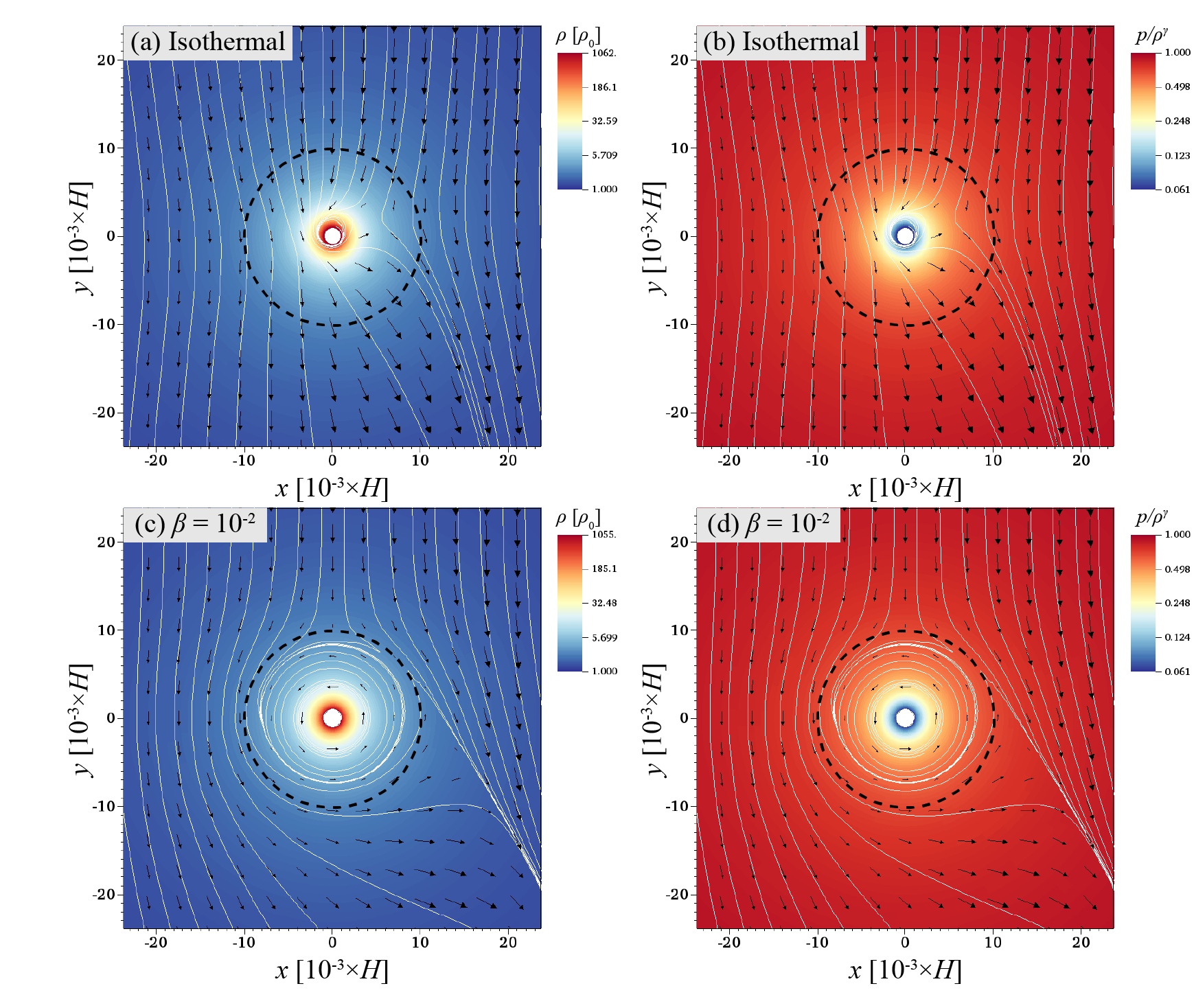}
    \caption{A comparison of flow patterns of (a,b) {\tt headw-is-m001} and (c,d) {\tt headw-B2-m001} (shear flow with headwind). The flow pattern is represented by arrows and streamlines. \textcolor{black}{The length of the arrows scales to the speed.} Colour contour represents (a,c) density and (b,d) $p/\rho^\gamma$ (entropy), respectively. A dashed circle denotes the Bondi radius.}
    \label{fig:head}
\end{figure*}

\begin{figure*}
    \centering
    \includegraphics[width=\linewidth]{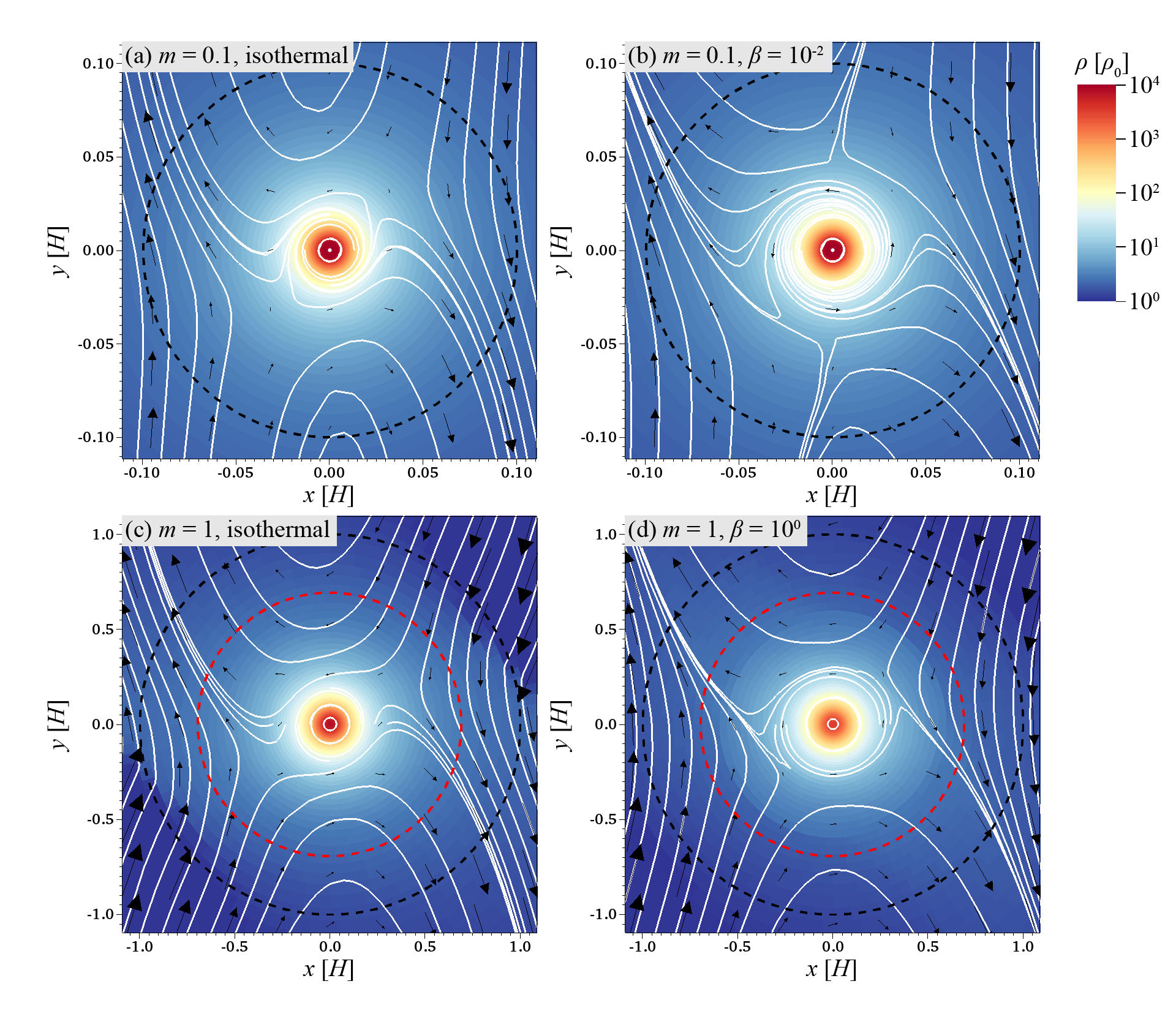}
    \caption{ \textcolor{black}{A comparison of flow patterns in the midplane: (a) {\tt shear-is-m01} at $t=15$, (b) {\tt shear-B2-m01} at $t=15$, (c) {\tt shear-is-m1} at $t=30$, and (d) {\tt shear-B0-m1} at $t=30$. The flow pattern is represented by arrows and streamlines. The length of the arrows scales to the speed, but the scale is different for each figure. The density is shown by colour contour. Dashed circles denote the Bondi (black) and Hill (red) radii.}}
    \label{fig:mass}
\end{figure*}

\begin{figure}
    \centering
    \includegraphics[width=\linewidth]{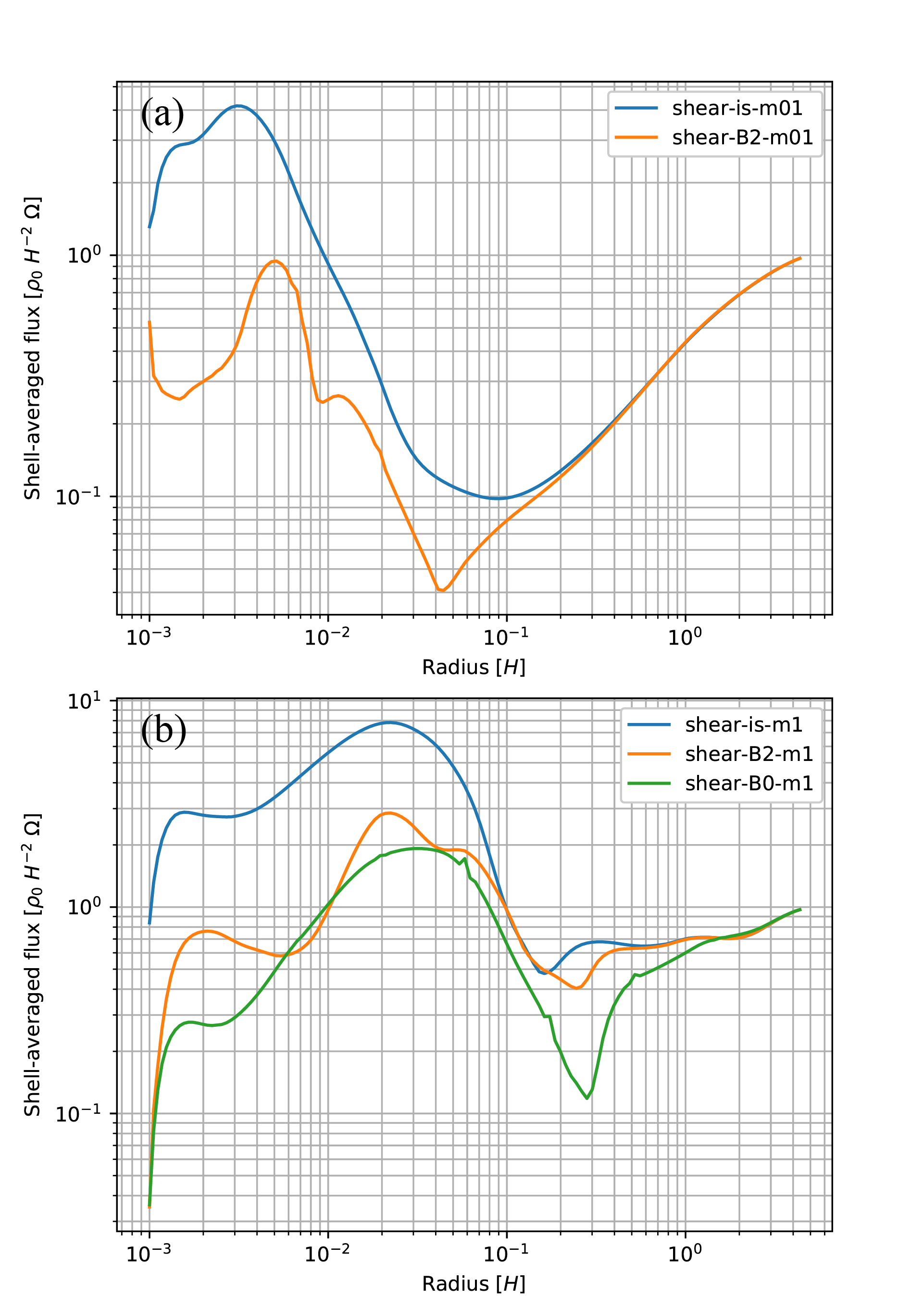}
    \caption{\textcolor{black}{Shell-averaged recycling flux (Equation \ref{eq:flux}). (a) A comparison between {\tt shear-is-m01} and {\tt shear-B2-m01} at $t=15$. (b) A comparison between {\tt shear-is-m1}, {\tt shear-B2-m1}, and {\tt shear-B0-m1} at $t=30$.}}
    \label{fig:r-flux-mass}
\end{figure}

\begin{figure}
    \centering
    \includegraphics[width=\linewidth]{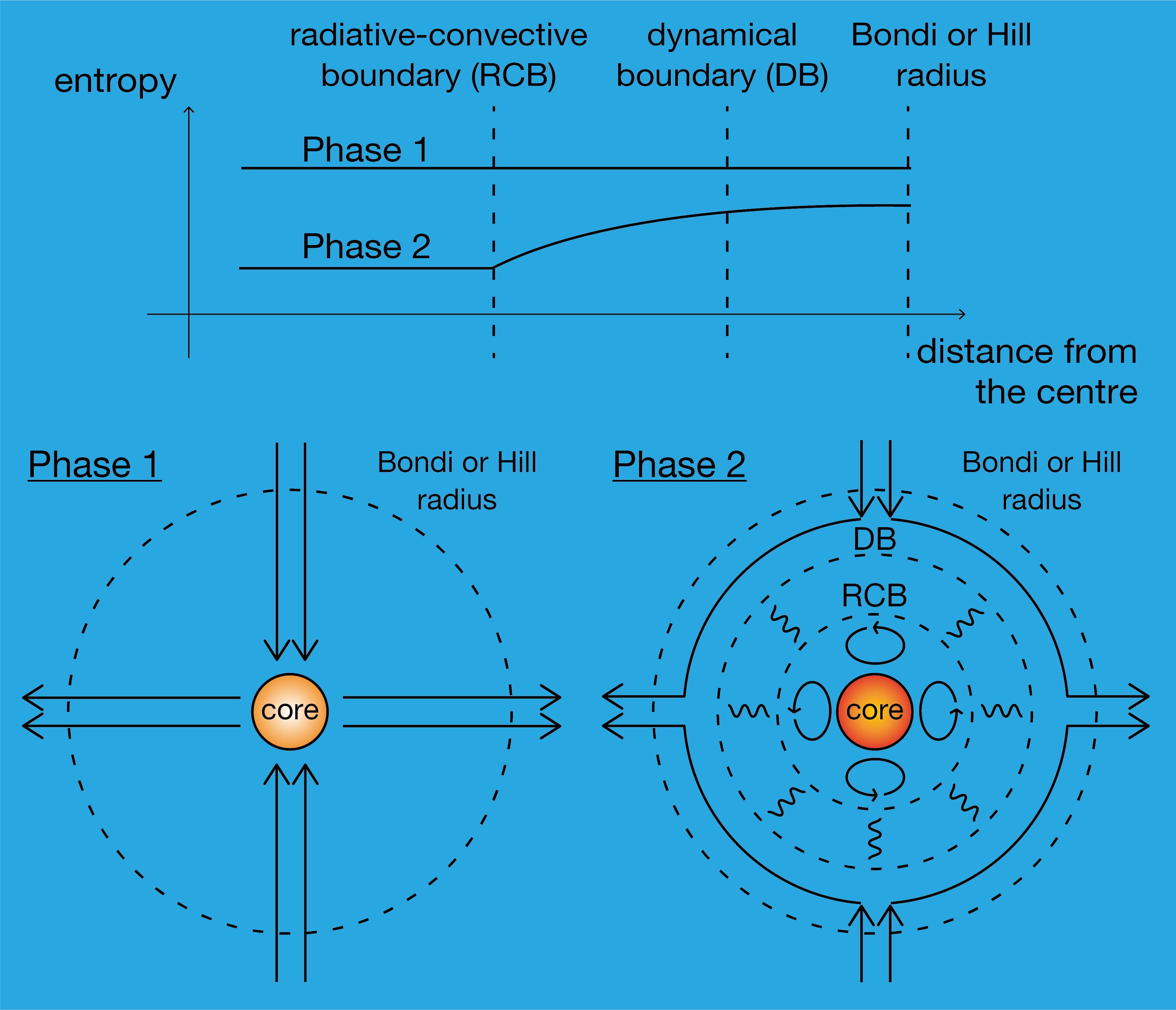}
    \caption{A schematic view of planetary envelopes embedded in a protoplanetary disc proposed in this study. \textcolor{black}{Phase 1}: When \textcolor{black}{the accretion rate of solid materials or luminosity of the core was high}, the envelope was fully convective. Complete recycling was possible at the time. Case 2: When the envelope started \textcolor{black}{cooling}, the radiative layer appeared and the buoyancy barrier began to prevent the disc-gas flow from recycling the envelope. The innermost envelope was still convective.}
    \label{fig:evolution}
\end{figure}

\subsection{A comparison between isothermal and non-isothermal simulations}

Our model showed differences in flow patterns of recycling between isothermal and non-isothermal cases.
We compared the time sequence in isothermal and non-isothermal cases in Figures \ref{fig:time_iso} \textcolor{black}{and \ref{fig:time_B2}} \textcolor{black}{for the nominal mass case ($m = 0.01$)}.
Gas started to accrete onto a planetary core as the gravity was gradually inserted with $t_{\rm inj}=0.5$ (Figures \ref{fig:time_iso}a and \textcolor{black}{\ref{fig:time_B2}a}).
At $t=2$, the density has reached its \textcolor{black}{steady-state} value (Figures \ref{fig:time_iso}b and \textcolor{black}{\ref{fig:time_B2}b}).
It means that a hydrostatic atmosphere has been built at this point.
\textcolor{black}{To the first approximation, a comparison between $t=10$ (Figures \ref{fig:time_iso}c and \ref{fig:time_B2}c) and $t=15$ (Figures \ref{fig:time_iso}d and \ref{fig:time_B2}d) shows that the velocity field has reached a steady state in terms of its basic properties (the presence/absence of atmospheric recycling; see following paragraphs), though the streamlines show some time-varying features.
The time-varying velocity field has also been reported in \citet{Ormel+2015}.}

The isothermal simulation \textcolor{black}{(Figure \ref{fig:time_iso})} showed outflow leaving the planet in the midplane as part of the recycling flow, which entered the Bondi sphere from the polar regions as shown by \citet{Ormel+2015}.  
In contrast, the non-isothermal run \textcolor{black}{(Figure \ref{fig:time_B2})} showed more circular streamlines around the planetary core within $r<\sim0.5\ R_{\rm Bondi}$ (remember that the dimensionless planetary mass $m$ is the measure of the Bondi radius), meaning that the recycling of envelope gas is limited and envelope gas is isolated in this region.
We observed streamlines heading outwards outside the isolated envelope but still within the Bondi sphere. 
The difference in flow patterns has been maintained until the end of the calculations (\textcolor{black}{$t=15$, Figures \ref{fig:time_iso}d and \ref{fig:time_B2}d}).

The non-isothermal simulations showed recycling flows similar to what has been reported in the isothermal case \citep{Ormel+2015}, but the recycling was observed only outside the isolated inner envelope.
We showed 3D streamlines in Figure \ref{fig:3D}a.
Fluid descended in polar regions as it circulated outside the isolated inner envelope.
Once the descending gas reached the midplane, it flowed outwards in the midplane.
In contrast, the isothermal simulations showed the recycling inflow and outflow being connected to the planetary core as reported previously \citep{Ormel+2015}.

\subsection{The dependence on the cooling time-scale} \label{subsec:dependence}

The size of the isolated inner envelope where we observed circular streamlines is positively correlated to the cooling time-scale $\beta$.
We compared the midplane flow patterns \textcolor{black}{ of the isothermal ($\beta \to 0$), $\beta = 10^{-4}$, $\beta = 10^{-3}$, $\beta = 10^{-2}$, and adiabatic ($\beta \to \infty$) simulations} at the same time well after the hydrostatic structure was developed ($t=7$) in Figure \ref{fig:comp_non-iso}.
Similar to the isothermal run (Figure \ref{fig:comp_non-iso}a), 
the outflow streamlines were launched from the vicinity of the core in the case where $\beta=10^{-4}$ was assumed (Figure \ref{fig:comp_non-iso}b).
We note that the flow patterns were not completely the same in the two cases.
The longer the cooling time-scale $\beta$ we assumed, the larger \textcolor{black}{region of circular streamlines} we observed (Figures \ref{fig:comp_non-iso}b-d).
\textcolor{black}{There exists a dynamical boundary: the inner envelope is isolated from the shear and horseshoe flow of disc gas (Figure \ref{fig:comp_non-iso}e).
As in the $\beta = 10^{-2}$ case (Figure \ref{fig:3D}a), the recycling flow was observed outside the isolated inner envelope (but still within the Bondi sphere) in all cases.}

\textcolor{black}{
The isolated inner envelope disappears in the slow cooling limit ($\beta \to \infty$). 
The adiabatic simulation did not show the dynamical boundary between the envelope and disc gas in its stream lines (Figure \ref{fig:comp_non-iso}f).
The absence of isolated envelope agrees with the result of the isothermal case, though the adiabatic case showed a rather time-varying flow pattern.
This is possibly because the smaller density of the envelope in the adiabatic simulation allowed the disc gas flow to perturb the envelope more easily.
As we will explain in subsection \ref{subsec:buoyancy}, a comparison of these simulations suggests that the presence or absence of the dynamical boundary is controlled by the entropy difference between the descending inflow gas and the envelope, which exists only in the non-isothermal (finite $\beta$) simulations.
We note that the nature of our adiabatic simulation is consistent with the previous study \citep{Popovas+2018}.
}

In contrast to the topology of the flow, the density profiles were almost the same in the simulations assuming different values of $\beta$ \textcolor{black}{except for the adiabatic case}. 
Because the flow was subsonic, the pressure structure was determined to be in hydrostatic equilibrium with the gravity of the planet. 
Temperature was nearly uniform due to the efficient radiative cooling.
Therefore, the density profiles did not depend on $\beta$.

\textcolor{black}{
In order to measure the efficiency of atmospheric recycling more quantitatively, Figure \ref{fig:r-flux} shows a comparison of the shell-averaged recycling and perturbation flux, which is defined by,
\begin{equation}
    \overline{F_{\rm out}}(r) \equiv \frac{\int \rho \cdot |v_r| \cdot r^2 d\Omega}{4\pi r^2}, \label{eq:flux}
\end{equation}
where the integral denotes a surface integral over the shell. 
Because the flux includes both actual recycling and perturbation within the bound atmosphere, the value represents the upper limit of the recycling flux.
Initially ($t<2$), the planet accreted the disc gas until the hydrostatic equilibrium was established, so that the flux within the Bondi sphere ($r < 10^{-2}$) was high due to the large influx (Figure \ref{fig:r-flux}a).
After the accretion period, the flux reached a nearly steady-state value. 
The larger $\beta$ we assumed, the smaller the recycling flux we obtained within the Bondi sphere ($r < 10^{-2}$, Figure \ref{fig:r-flux}b).
The flux in the case of $\beta = 10^{-2}$ is smaller than that in the isothermal simulations by one or two orders of magnitude.
It means that the envelope became more stable against the recycling as we assumed slower (inefficient) cooling of the descending disc gas (see subsection \ref{subsec:buoyancy}).
The change in the position of dynamical boundary between the atmosphere and disc gas found in Figure \ref{fig:comp_non-iso} is also visible in Figure \ref{fig:r-flux}b: the position where the flux has a minimum moved to outer regions as we assumed larger $\beta$.
Because the disc gas far from a planet has shear flow in all cases, the flux values of isothermal and non-isothermal simulations outside the Bondi sphere matched each other.
}

\subsection{The buoyancy barrier to prevent recycling}
\label{subsec:buoyancy}

The emergence of an isolated inner envelope around a planet in the non-thermal simulations is due to {\it the buoyancy barrier} to descending gas caused by the entropy gradient.
Figure \ref{fig:entropy} showed the entropy structure around a planet in the models shown in Figure \ref{fig:comp_non-iso}.
The entropy structure was \textcolor{black}{similar for isothermal and non-isothermal simulations (except for the adiabatic case)}: there is a low-entropy region surrounding the planet, which is spherically symmetric, while the entropy is uniform in the outer region (Figures \ref{fig:entropy}a-e).
The pressure structure was determined to be in hydrostatic equilibrium with the gravity of the planet. 
The radiative cooling made the temperature nearly uniform and equal to that of the background shear flow.
As a result, the entropy was lower around the planet.
\textcolor{black}{In contrast, the adiabatic simulation has an isentropic structure (Figure \ref{fig:entropy}f).
A low-entropy shell exists around the inner boundary because of an artificial cooling term introduced to the innermost cell of the grid (subsection \ref{subsec:settings}), but its influence was limited to few cells in the vicinity of the inner boundary.}

The positive entropy gradient inhibits the descending gas from reaching deeper regions in the envelope \textcolor{black}{in the non-isothermal cases}.
Let us suppose a parcel of descending gas.
Because the flow is subsonic, the pressure of the gas parcel is equal to that of the surrounding gas.
In contrast, the temperature would increase as the gas descends because of adiabatic compression with some radiative loss.
Consequently, the parcel has a higher temperature compared to that of its surroundings and so the density is lower in the parcel. 
The density difference exerts buoyancy, which prevents further descent.
The buoyancy caused by the entropy gradient is well-known as one of the major factors governing stellar and planetary structures \citep[e.g.][]{Kippenhahn+2012}.

\textcolor{black}{
The isothermal and adiabatic simulations overestimate the efficiency of atmospheric recycling.
Because the temperature of descending disc gas is immediately equilibrated with that of the surroundings, no buoyant force is exerted on the gas in the isothermal limit.
Thus, efficient atmospheric recycling is allowed.
Whereas in the adiabatic limit, the descending gas has the same entropy as its surroundings.
Again, the absence of buoyancy allows efficient recycling.
}

The importance of the buoyancy barrier for the emergence of an isolated envelope can be demonstrated by simplified simulations.
We compared the flow topology in the isothermal and non-isothermal simulations again in Figure \ref{fig:headonly}, but here we removed the complexity -- a planet is embedded in uniform headwind without a shear component and inertial force: the Coriolis, tidal, and global pressure forces.
The density structure was the same because it is determined by the hydrostatic equilibrium with the planet gravity (Figures \ref{fig:headonly}a and \ref{fig:headonly}c).
The temperature was nearly uniform, and, consequently, the entropy structure is also the same and spherically symmetric (Figures \ref{fig:headonly}b and \ref{fig:headonly}d).
The flow patterns were different between the isothermal and non-isothermal runs.
The isothermal run showed that the flow focuses towards the gravitating body because of the mass conservation: as the density increases around the planet, the speed of the flow has to decrease in order to conserve the mass flux, leading to the flow converging towards the centre (Figures \ref{fig:headonly}a and \ref{fig:headonly}b).
The convergent flow in the isothermal case has been reported previously \citep{Lee+Stahler2011,Ormel2013}.
On the contrary, the headwind diverged in front of the planet and an isolated envelope formed there in the non-isothermal case (Figures \ref{fig:headonly}c and \ref{fig:headonly}d).
As we have omitted all the complexity in these runs, the emergence of the isolated envelope can be regarded as a consequence of the buoyancy barrier discussed above.
\textcolor{black}{This interpretation will hold in more realistic simulations performing radiative transfer calculations (see subsection \ref{subsec:previous}).}

\subsection{The cases with sub-Keplerian disc}

Planets are exposed to headwind when disc gas rotates sub-Keplerian.
The buoyancy barrier prevented the headwind from penetrating the planetary envelope entirely, even in these cases where both the headwind ($M_{\rm hw}=0.1$, Table \ref{tab:models}) and shear flow were taken into account.
The flow patterns in the isothermal and non-isothermal simulations were compared in Figure \ref{fig:head}.
The density and entropy structures were again the same because of the hydrostatic equilibrium with the planetary gravity and the efficient radiative cooling.
The topography of the flow differs between the isothermal and non-isothermal simulations in the headwind case as well as the shear-flow case (Figure \ref{fig:comp_non-iso}).
Whereas the isothermal run showed that the headwind penetrated deeply into the Bondi sphere (Figures \ref{fig:head}a and \ref{fig:head}b), 
the non-isothermal run showed an isolated envelope around the planet where the atmospheric recycling was limited (Figures \ref{fig:head}c and \ref{fig:head}d). 
This is because of the buoyancy preventing the headwind from penetrating the envelope.

The non-isothermal simulation with the headwind shares a similarity with the isothermal one when it comes to the flow outside the isolated envelope.
We showed 3D streamlines in Figure \ref{fig:3D}b.
The headwind blowing against the envelope in the mid plane slid down along the boundary of the isolated envelope and left above the midplane.
A similar flow pattern has been reported in the isothermal simulation \citep{Ormel+2015}, while the headwind has reached the planetary core in that case.

\subsection{The dependence on planetary mass}

\textcolor{black}{
The buoyancy barrier prevents the atmospheric recycling regardless of the planetary mass.
Figure \ref{fig:mass} compares the midplane flow patterns of the isothermal and non-isothermal simulations for $m = 0.1$ and $m = 1$.
The size of the circular-streamline region is larger in the non-isothermal simulations than in the isothermal simulations, as with the $m = 0.01$ cases.
The recycling and perturbation flux (Equation \ref{eq:flux}) was smaller in the non-isothermal simulations than in the isothermal simulations by an order of magnitude (Figure \ref{fig:r-flux-mass}).
It suggests that the buoyancy barrier induced by an entropy gradient prevented the inflow from descending onto a planet in the non-isothermal simulations.
We note that those higher-mass simulations have not reached the steady state at the time of termination: the density in the vicinity of the planet has still been increasing gradually.
Their behaviour in more long-term simulations will be addressed in our future work.}

\textcolor{black}{
A difference from the lower mass simulation (Figure \ref{fig:comp_non-iso}a) is that the streamlines in the vicinity of the planet ($\sim 0.2\ R_{\rm Bondi}$) is circular even in the case of the isothermal simulations (Figures \ref{fig:mass}a and \ref{fig:mass}c).
Because higher-mass planets accrete more disc gas in a more 2D-fashion, they obtain more angular momentum than lower-mass planets \citep[more precisely, because of the vortensity conservation in 2D flow, ][]{Ormel+2015a}.
As a consequence, higher-mass planets have more circular streamlines even in the isothermal simulations.
This picture is consistent with an inference that planets form circumplanetary discs in the higher-mass limit \citep[e.g.,][]{Tanigawa+2012}.
}

\section{Discussion} \label{sec:discussion}

\subsection{Interpreting the results of previous studies}
\label{subsec:previous}

We showed that the buoyancy barrier prevents the disc-gas inflow from recycling the gas in the planetary envelope by comparing isothermal and non-isothermal calculations.
Whereas the radiative cooling was implemented by the simple $\beta$ cooling approximation in our model, the results are useful to understand the difference between previous 3D simulations, some of which have carried out more sophisticated, radiative-transfer calculations.

Isothermal simulations of flow past a planet in a protoplanetary disc have pointed to complete recycling of the envelope -- no closed streamlines have been found around a planet.
\citet{Ormel+2015} simulated inviscid isothermal flow around an embedded planet \textcolor{black}{(m = 0.01)} in a local frame.
They found that no clear boundary demarcates bound atmospheric gas from disc material: gas enters the Bondi sphere at high latitudes and leaves through the midplane regions or, vice versa when the rotation of disc gas is sub-Keplerian.
\citet{Fung+2015} presented global simulations of the isothermal flow around an embedded planet \textcolor{black}{(m = 0.56)} with viscosity.
They found a \textquotedblleft transient" horseshoe flow along which high altitude gas descends rapidly into the planet's Bondi sphere, performs one horseshoe turn, and exits the Bondi sphere radially in the midplane.
The flow prevented the planet from sustaining a hydrostatic envelope, that is, the gas is recycling.
We note that they found a bound gas in the deep part of the envelope within $\sim 1.5$ times the size of the planetary core.
Because the bound region was resolved by only 3 grids, 
they argued that it is uncertain how much gas is truly bound to the planet.

\textcolor{black}{
Adiabatic simulations also found no dynamical boundary between bound atmosphere and disc gas.
\citet{Popovas+2018} simulated inviscid adiabatic flow around an embedded planet in a local frame.
Their numerical set up assumed Mars- to Earth-sized planets at $1-1.5$ AU, which corresponds to $m = 0.011, 0.037$ and $0.070$ in the dimensionless mass used in our study.
The streamlines within the Bondi sphere were non-circular and connected to disc gas. 
}

In contrast, \textcolor{black}{radiation-hydrodynamics} simulations have reported limited recycling -- an inner part of the envelope is isolated from the recycling flow of disc gas.
\citet{DAngelo+Bodenheimer2013} performed global 3D radiation-hydrodynamics calculations of protoplanetary discs in which a young planetary core is located with explicit viscosity.
Realistic equation of state and gas opacity were used and the energy released by the accretion of planetesimals was considered in their simulations.
\textcolor{black}{They assumed 5--15 ${\rm M_{\earth}}$ planets located at 5--10 AU, which corresponds to $m = 0.080$--0.25.}
They found that the interface between the material orbiting the planet and material orbiting the star lies at $\sim0.4\ R_{\rm Bondi}$.
The global radiation-hydrodynamics simulations of \citet{Lambrechts+Lega2017} also reported that the interface is at \textcolor{black}{$\sim0.3$} times the Hill radius of the planet.
\textcolor{black}{They assumed 5 $M_{\earth}$ and larger planets at 5 AU, which corresponds to $m > 0.3$.}
\citet{Cimerman+2017} conducted inviscid radiation-hydrodynamics calculations of a flow around a planet in a local frame.
\textcolor{black}{The assumed planetary masses were $m = 0.04$, $0.38$, $0.75$, and $1.9$.}
They identified an \textquotedblleft inner core" of the envelope -- $\sim 0.25\ R_{\rm Bondi}$ in size -- where streamlines are more circular and entropy is much lower than in the outer envelope. 

To summarise, complete recycling has been found in isothermal \textcolor{black}{and adiabatic} calculations; a limited recycling has been observed in non-isothermal calculations, regardless of many differences in these previous studies (local/global, viscid/inviscid, equation of state, opacity, planetesimal accretion, and planetary masses).
The clear difference between previous isothermal and non-isothermal studies is consistent with our explanation that the high-entropy disc-gas is prevented from descending into the low-entropy envelope by the buoyant force in the non-isothermal cases.

We note that the entropy gradient has also been proposed to influence the gas accreting onto young stars: high-entropy gas accreting onto stars cannot descend into the stellar interior with low entropy \citep{Geroux+2016,Kunitomo+2017}.
These studies seem to be consistent with our idea that high-entropy disc gas is prevented from descending into the low-entropy envelope in the case of gas accretion onto planets.

\subsection{The effect of recycling on gas accretion} \label{subsec:accretion}

The buoyancy barrier suppresses the atmospheric recycling, which might allow the cooling and accretion of the envelopes of super-Earths.
\textcolor{black}{\citet{Malygin+2017} estimated the radiative relaxation time-scale of linear temperature perturbations in protoplanetary discs.
\textcolor{black}{The relaxation time-scale $t_{\rm relax}$ in the optically thick regime was estimated by, 
\begin{equation}
t_{\rm relax} \sim \frac{l^2}{D} \sim \frac{l^2 \kappa_{\rm R} \rho}{\chi c}, \label{eq:relax}
\end{equation}
with $l$ being the length scale of perturbation, $D$ the diffusion coefficient, $\kappa_{\rm R}$ the Rosseland mean opacity, $c$ the speed of light, and $\chi$ the flux limiter \citep{Malygin+2017}.}
A parcel of descending gas of inflow into the Bondi sphere of a planet (subsection \ref{subsec:buoyancy}) would be heated by adiabatic compression.
The typical scale of this temperature increase can be given by the Bondi radius: $R_{\rm Bondi} = H \times m$.
The estimated relaxation time is $\sim 10^{-2}\ \Omega^{-1}$ for the perturbation length scale of $\lambda \sim 0.01\ H$ at 1 AU \citep[][in their nominal model without dust depletion]{Malygin+2017}, which corresponds to the size of the Bondi radius of a $m=0.01$ planet.
As we presented in subsection \ref{subsec:dependence}, the simulation with this relaxation time ($\beta = 10^{-2}$) showed the emergence of an isolated inner envelope.
The relaxation time becomes longer for larger scales (namely, higher-mass planets) and smaller orbital radius \citep{Malygin+2017}.
Because the atmospheric recycling becomes less efficient for larger $\beta$ (subsection \ref{subsec:dependence}), our results suggest that runaway gas accretion onto super-Earths may be inevitable once the envelopes have started cooling.}
Fluid close to the core is bound to the planet and modified 1D prescriptions may be suitable, where the cooling controls the accretion \citep{DAngelo+Bodenheimer2013,Lambrechts+Lega2017}.
\textcolor{black}{We note that the relaxation time depends on the dust opacity: the dust depletion would decrease the relaxation time \citep{Malygin+2017} and, consequently, the size of the isolated inner envelope, although dust depletion would also decrease the time-scale of envelope cooling and runaway accretion \citep{Lee+2014}.}

\textcolor{black}{There is a caveat in our analysis by using the $\beta$ approximation: the presence of a planet is not a linear perturbation to the disc in reality.
Because the gas density in Equation \ref{eq:relax} is largely modified in the vicinity of the planet, the linear approximation is no longer valid in this region.
However, our simulation showed that the descending disc gas does not reach the deep envelope and stays in the outer region where gas density is not significantly modified by the planet.
Thus, we argue that the behaviour of the recycling flow was captured in our simulations, 
though a comparison to the radiation-hydrodynamics simulations (subsection \ref{subsec:previous}) was necessary to confirm the validity. 
}

The presence of the isolated envelope does not necessarily mean that the recycling mechanism cannot explain the ubiquity of super-Earths.
\citet{Cimerman+2017} has found a low-entropy outflow leaving the planet from the inner core of the envelope.
Advection flow of gas on the low-entropy core of the envelope seems to have removed a fraction of the gas.
They stated that studies at higher resolutions are needed to assess whether this region can become hydrodynamically isolated on long time-scales.

We propose that the structures of planetary envelopes embedded in a protoplanetary disc evolved as follows (Figure \ref{fig:evolution}).
When \textcolor{black}{the accretion rate of solid materials or luminosity of the core was high} (\textcolor{black}{Phase} 1), the envelope was fully convective \textcolor{black}{\citep{Rafikov2006,Lambrechts+Lega2017,Popovas+2018}} and had an \textcolor{black}{isentropic} profile. 
Complete recycling would have been allowed because there was no buoyant force on the descending disc gas \textcolor{black}{(Figures \ref{fig:comp_non-iso}}f and \ref{fig:entropy}f, subsection \ref{subsec:buoyancy}).
\textcolor{black}{We note that Phase 1 may not be realised for typical values of the solid accretion rate and opacity \citep[e.g.,][]{Ikoma+2000,Lambrechts+2014,Lambrechts+Lega2017,Lee+2017}.}
Once the envelope started to be cooled (\textcolor{black}{Phase} 2), a radiative layer characterised by a positive entropy gradient appeared.
The buoyancy barrier would begin to prevent the disc gas from descending into the envelope \textcolor{black}{(subsection \ref{subsec:buoyancy})} so that the envelope would have three layers: convective, radiative, and recycling.
The three layer structure has been proposed by \citet{Lambrechts+Lega2017}.
The radiative-convective boundary is set by the convective stability:  
the local entropy gradient determines the direction of buoyancy force and, consequently, the stability. 
This is the same mechanism with the buoyancy barrier which was proposed to set the boundary for the recycling flow in this study.
\textcolor{black}{The innermost convective layer was missing in our models (Figure \ref{fig:comp_non-iso}) because we neglected heat sources: the accretion of solid materials, luminosity of the core, and the release of energy from gas.
However, the convective layer is likely to exist in reality.}

\subsection{Implications for the formation of super-Earths}

The local flow past a planet investigated in this study may help to explain the formation of super-Earths by delaying the  growth of their solid cores.

In the pebble accretion theory \citep{Ormel+Klahr2010,Lambrechts+Johansen2012}, 
proto-cores can grow up to a mass heavy enough to carve a gap in the pebble disc, which is given by \citep{Lambrechts+2014},
\begin{equation}
M_{\rm iso}^{\rm pebble} \sim 20 \biggl( \frac{H/a}{0.05} \biggr)^3 M_{\earth}.
\end{equation}
This mass is large enough to allow disc gas to accrete in a runaway fashion within the lifetime of a protoplanetary disc \citep{Lee+2014}.

Outflow in the midplane region may reduce the pebble/dust accretion rate onto a planetary core and, \textcolor{black}{consequently}, delay the core growth.
\textcolor{black}{The flow induced by a planet has been shown to reduce the accretion rate of small dust particles as they closely follow streamlines, while pebble-sized particles are only weakly affected \citep{Ormel2013,Popovas+2018}.}

\citet{Alibert2017} has proposed that the outflow as \textquotedblleft advection wind" combined with the vaporisation of pebbles in the envelopes prevents the growth of the planets at masses smaller or similar to Earth mass.
Our results suggested that the advection wind might not reach the inner envelope where pebbles vaporise.
In addition, the advection wind containing the vaporised pebbles as heavy elements would have a higher density than that of surrounding gas. 
The buoyancy barrier might prevent the advection wind from ascending from the planet's gravity potential, \textcolor{black}{though analysis of convective stability considering both composition and temperature gradients \citep{Kurokawa+Inutsuka2015} is needed to understand the consequences.}

In either case, further studies on the interaction of the planet-induced wind with pebbles/dusts are needed to understand the consequence on the formation scenarios of super-Earths.

\section{Conclusion} \label{sec:conclusion}

The ubiquity of super-Earths in extrasolar planetary systems poses a problem for the theory of planet formation as they are expected to become gas giants through the runaway accretion of disc gas on a short time-scale.
\citet{Ormel+2015} has proposed that rapid recycling of the envelope gas of planets embedded in a protoplanetary disc would delay the cooling and following accretion of disc gas.
However, the topography of the recycling flow has been shown to differ between previous simulations of the flow past a planet in a disc.

We conducted a detailed comparison between isothermal and non-isothermal 3D hydrodynamical simulations of the gas flow past a planet to investigate the effect of radiative cooling on the recycling flow pattern.
Radiative cooling was implemented by the $\beta$ cooling model.

We observed the pattern of recycling flow: fluid entered the Bondi sphere at high latitudes and leaves through the midplane regions.
The topography of inflow and outflow is the other way around when disc gas rotates in sub-Keplerian. 
Whereas the isothermal simulations showed \textcolor{black}{efficient} recycling where streamlines are connected to \textcolor{black}{the inner part of the envelope gas}, 
the non-isothermal runs showed limited recycling: at the inner part of the planetary envelope, streamlines are circular.
We demonstrated that the buoyancy force induced by the entropy difference between the atmosphere (low entropy) and disc gas (high entropy) suppresses the recycling.
Though our models adopted the simple $\beta$ cooling, the mechanism of the buoyancy barrier is useful to understand the difference between the previous studies.

Our results suggested that, once the atmosphere starts cooling, the buoyancy barrier prevents the high-entropy disc gas from intruding the cooled atmosphere, which may lead further cooling of the atmosphere and runaway gas accretion onto the core.
Nevertheless, the interaction of the recycling flow with accreting \textcolor{black}{solid materials} may delay the growth of super-Earth cores and help to explain the ubiquity of super-Earths in exoplanetary systems.

\section*{Acknowledgements}

\textcolor{black}{We would like to thank the reviewer, Michiel Lambrechts for thoughtful and helpful comments that have substantially improved the quality of this manuscript.}
We thank Athena++ developers: James M. Stone, Kengo Tomida, and Christopher White.
HK particularly acknowledges assistance by Kengo Tomida to start using the code.
This study has greatly benefited from fruitful discussion with Chris W. Ormel, Shu-ichiro Inutsuka, Masanobu Kunitomo, \textcolor{black}{Shigeru Ida, Ayumu Kuwahara, Takayuki Saito, and Masahiro Ikoma}.
HK was supported by JSPS KAKENHI Grant number 15J09448 \textcolor{black}{and 18K13602}.
TT was supported by JSPS KAKENHI Grant number 26800229 and 15H02065.
Numerical computations were in part carried out on Cray XC30 at the Center for Computational Astrophysics, National Astronomical Observatory of Japan.








\appendix


\bsp	
\label{lastpage}
\end{document}